\documentclass[]{spie}  
 
\usepackage{amsmath,amsfonts,amssymb}
\usepackage{graphicx}
\usepackage[colorlinks=true, allcolors=blue]{hyperref}
\usepackage{float}

\title{Output-Stage Design Optimization for High-Sensitivity SiSeRO CCDs and SiSeRO Active Pixel Sensors
}

\author[a]{Tanmoy Chattopadhyay}
\author[a]{Sven Herrmann}
\author[b]{Kevan Donlon}
\author[b]{Ilya Prigozhin}
\author[a]{Peter Orel}
\author[a,c,d]{Steven W. Allen}
\author[e]{Marshall W. Bautz}
\author[b]{Michael Cooper}
\author[e]{Catherine E. Grant}
\author[e]{Beverly LaMarr}
\author[b]{Christopher Leitz}
\author[e]{Eric D. Miller}
\author[a]{R. Glenn Morris}
\author[a,c]{Abigail Y. Pan}
\author[a]{Tonya L. Peshel}
\author[a]{Artem Poliszczuk}
\author[e]{Gregory Prigozhin}
\author[a,c]{Haley R. Stueber}
\author[b]{Keith Warner}

\affil[a]{Kavli Institute of Astrophysics and Cosmology, Stanford University, 452 Lomita Mall, Stanford, CA 94305, USA}
\affil[b]{MIT Lincoln Laboratory, Lexington, MA, USA}
\affil[c]{Department of Physics, Stanford University, 382 Via Pueblo Mall, Stanford CA 94305, USA}
\affil[d]{SLAC National Accelerator Laboratory, 2575 Sand Hill Road, Menlo Park, CA 94025, USA}
\affil[e]{Kavli Institute for Astrophysics and Space Research, Massachusetts Institute of Technology, Cambridge, MA, USA}

\authorinfo{Further author information: Send correspondence to T. Chattopadhyay\\T. Chattopadhyay: E-mail: tanmoyc@stanford.edu}

\begin{document} 
\maketitle


\begin{abstract}
The Single electron Sensitive Read Out (SiSeRO) technology is a new device class designed to support the needs of future X-ray and optical astronomical telescopes that will require fast, low-noise, megapixel spectro-imagers. Developed at MIT Lincoln Laboratory, in collaboration with Stanford University and MIT, the first generation SiSeRO-CCD (charge-coupled device) prototypes achieved a charge/current conversion gain of 700$-$800 pA per electron, an equivalent noise charge (ENC) of around 3.5 electrons root mean square (RMS), and a full width half maximum (FWHM) energy resolution of approximately 130 eV at 5.9 keV at a readout speed of 625 kpix/s. Utilizing Repetitive Non-Destructive Readout (RNDR), these same devices also demonstrated sub-electron noise performance (ENC$<$0.5 electrons RMS) at a readout speed of 10 kpix/s. We present the results of device simulations for next-generation SiSeRO CCD output stages that optimize the sensing transistor and its internal gate geometry to enhance noise and speed performance. Further, the goal is to develop a SiSeRO active pixel sensor (APS) that  combines the proven X-ray performance of CCDs with the architectural advantages of an APS. Enabling this requires substantial design updates, for example, incorporating two SiSeRO amplifiers side by side on each pixel and shuffling the charge between them to support RNDR. We discuss our device simulation framework and design parameter optimization in the first-generation SiSeRO devices.  
\end{abstract}




\section{INTRODUCTION}
\label{sec:intro}   
Future X-ray astronomy missions require detectors capable of simultaneously achieving high frame rates, low read noise, and large imaging formats. The advanced X-ray charge-coupled devices (CCDs) provide excellent noise and energy resolution at significantly faster readout speeds \cite{bautz18,bautz19,bautz20,chattopadhyay22_ccd,Bautzetal2022,bautz24_ccd_spie,miller2023_spie_axis,stueber2025_ccd,miller_axis_ccd_2025} than the conventional CCDs, but the frame rate is still limited by serial readout architectures. Complementary Metal Oxide Semiconductor (CMOS) based active pixel sensors offer high-speed operation but generally exhibit higher noise levels than CCD technologies like in the case of Hybrid CMOS detectors\cite{chattopadhyay18_HCDoverview,hull17} or lack in the energy bandwidth, e.g. the monolithic CMOS detectors\cite{kenter19_mcmos}.

The Single electron Sensitive Readout (SiSeRO\cite{chattopadhyay22_sisero,chattopadhyay_sisero_asic_2025,Donlonpie2024}) architecture, developed at MIT Lincoln Laboratory, is a novel charge detection technology for X-ray CCDs. In a SiSeRO device, signal electrons are collected in an internal gate located beneath a p-channel MOSFET output transistor. Variations in stored charge modulate the transistor drain current, enabling meaurement of charge. It also enables repeated measurements of the charge and sub-electron effective read noise through Repetitive Non-Destructive Readout (RNDR).

Current SiSeRO CCDs demonstrate read noise of approximately 3.5 $\mathrm{e}^{-}_{\mathrm{RMS}}$ at 625 kHz of readout speed\cite{chattopadhyay23_sisero,Chattopadhyayetal2022} and sub-electron noise performance using RNDR  \cite{chattopadhyay24_rnrdr,chattopadhyay24_rndr_spie}. However, device performance remains limited by output-stage gain and excess low-frequency noise. Improving these characteristics is critical for the development of next generation high sensitivity SiSeRO CCDs and for extending SiSeRO technology to large-format active pixel sensors (APS) suitable for future flagship missions such as Lynx\cite{gaskin19}.

In this work, we use Sentaurus TCAD simulations\cite{synopsys_sentaurus_tcad} to investigate the dependence of SiSeRO output-stage performance on key process and device parameters and identify optimization pathways for future detector generations.


\section{SiSeRO Output Stage and Simulation Methodology}
\subsection{SiSeRO output-stage concept}
The SiSeRO output stage consists of a p-type buried-channel MOSFET fabricated above an internal gate region. Signal electrons stored in the internal gate modify the channel potential and modulate the drain current. 
\begin{figure}[t!]
    \centering
   \includegraphics[trim={0.8cm 0.8cm 1cm 0.6cm}, clip=true,width=.8\linewidth]{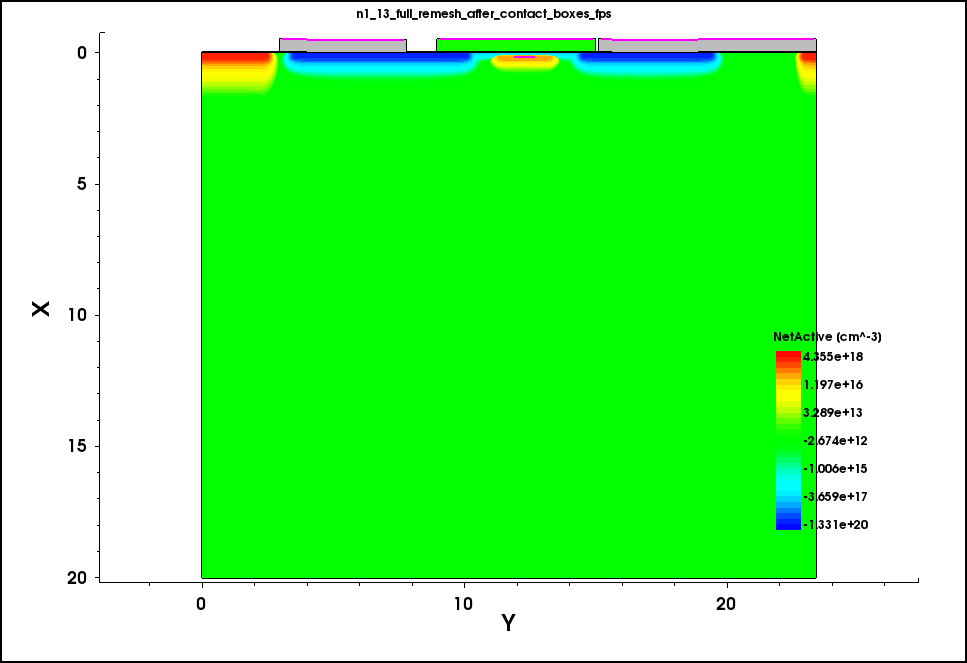}
    \caption{SiSeRO output stage in two-dimensional Sentaurus TCAD model. The model includes p type silicon substrate, a p-MOSFET transistor, source and drain implants, polysilicon gate and an internal gate beneath the p-MOSFET.}
    \label{fig:sim_sisero2d}
\end{figure}

A two-dimensional TCAD model of the output-stage transistor was developed in Synopsys Sentaurus as shown in Fig. \ref{fig:sim_sisero2d}. The structure was developed in \textit{sprocess} software of Sentaurus using the same fabrication process technology used to fabricate the first generation SiSeRO CCDs. The simulation geometry includes a p-substrate silicon of 5 k$\Omega$-cm resistivity, a p-type buried MOSFET channel, gate-stack consisting of 30 nm oxide and 20 nm nitride, polysilicon gate electrode, p+ source/drain implants, CCD n-channel, and internal gate trough structure. The internal gate structure consists of a 0.8 $\mu$m n-type trough deeper into the CCD buried channel and an additional 2 $\mu$m n-type sub-channel implant under the buried channel and the trough providing isolation between the p-channel of the MOSFET and the p-substrate. There are also two n+ implants (scuppers) on the source and drain side which helps to isolate the sense node from the substrate.

\subsection{Electrical simulations}
Current-voltage characteristics were simulated using thermodynamic transport models at temperature of 300 k including field-dependent mobility, quantum corrections, and Shockley-Read-Hall recombination in the \textit{sdevice} software of Sentaurus. All the device simulations were done with source and drain biased at 0 V and -3 V respectively, and gate voltage swept from +4 V (transistor off) to -3 V (saturation). The substrate was biased at 0 V for the simulations.
Transconductance was extracted from transfer characteristics of the device.
Detector gain was evaluated by introducing discrete charge in the internal gate and calculating the corresponding change in drain current. To introduce charge, we populate the internal gate with electron filled traps (therefore negatively charged) and make sure that the trap states do not alter during the simulation run (remains always filled). The total charge present in the internal gate is then computed by integrating the already existing electron density and the electron trapped charge density over the trough region.

\subsection{Noise simulations}
Noise simulations were performed using the Impedance Field Method (IFM\cite{dinh2012}). Thermal diffusion noise, interface trap noise, and bulk defect noise sources were included. Frequency-domain simulations (ac simulations) were carried out over the range relevant for SiSeRO readout applications (1 KHz to 10 MHz).

The investigated process parameters and their range are summarized in Table \ref{tab:optimization}.

\begin{table}[ht]
\caption{Summary of TCAD parameter optimization studies for the SiSeRO output-stage transistor.}
\label{tab:optimization}
\centering
\begin{tabular}{|l|c|c|}
\hline
\textbf{Parameter} &
\textbf{Range Studied} &
\textbf{Primary Metric} \\
\hline

MOSFET channel implant dose &
$5e12 - 8e12$ cm$^{-2}$ &
$g_m$, $G_q$ \\
\hline

Gate oxide thickness ($t_{ox}$) &
$10 - 30$ nm &
$g_m$, $G_q$ \\
\hline

Trough implant energy &
$125 - 200$ keV &
$G_q$ \\
\hline

Trough size ($W_{tr}$) &
$0.4 - 1.0$ $\mu$m &
$G_q$ \\
\hline

Trough location &
Source $\rightarrow$ Drain &
$G_q$ \\
\hline

Trough implant dose &
$4e11 - 1e13$ cm$^{-2}$ &
Charge confinement \\
\hline

Interface trap density ($D_{it}$) &
$1e12 - 1e14$ cm$^{-2}$eV$^{-1}$ &
Noise \\
\hline

Bulk defect density &
$1e14- 1e17$ cm$^{-3}$ &
Noise \\
\hline

Polysilicon gate doping &
$5e15 - 5e16$ cm$^{-2}$ &
Noise \\
\hline

\end{tabular}
\end{table}

\section{Optimization of Transconductance}
Transconductance ($g_m$) determines the sensitivity of the output transistor to changes in gate potential and directly impacts detector gain and noise performance and the bandwidth or speed of the detector.
Two primary parameters were investigated $-$
\begin{itemize}
    \item MOSFET buried-channel implant dose
    \item Gate oxide thickness
\end{itemize}

\begin{figure}[ht!]
    \centering
   \includegraphics[trim={0cm 0cm 0cm 1.4cm}, clip=true,height=5.3cm, keepaspectratio]{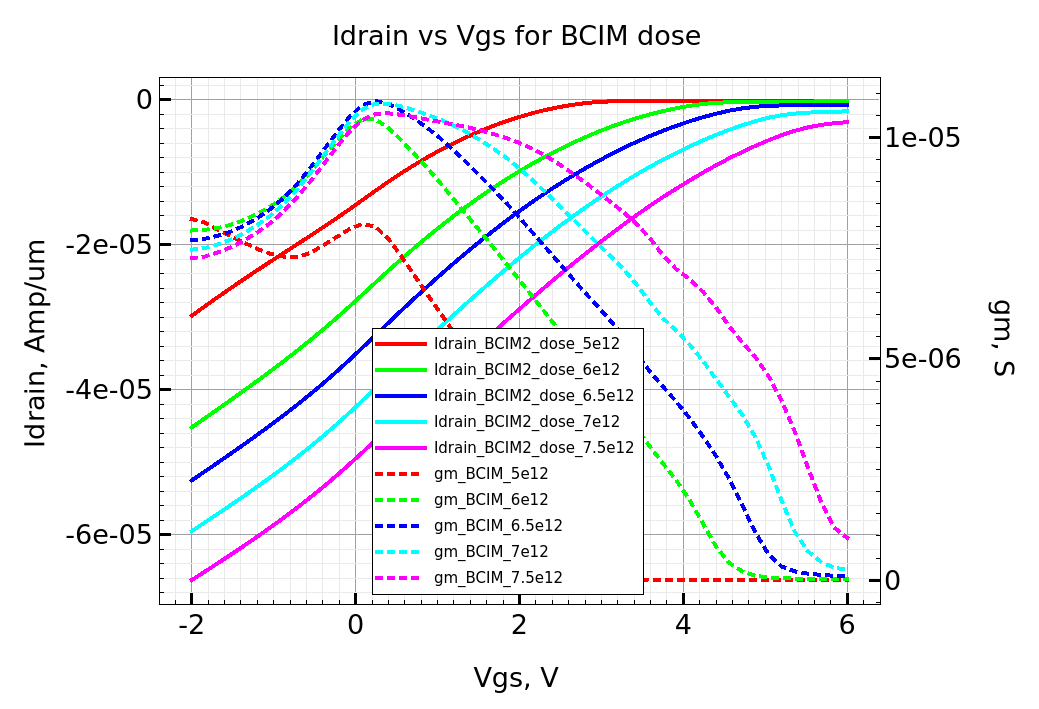}
   \includegraphics[trim={0cm 0cm 0cm 1.4cm}, clip=true,height=5.3cm, keepaspectratio]{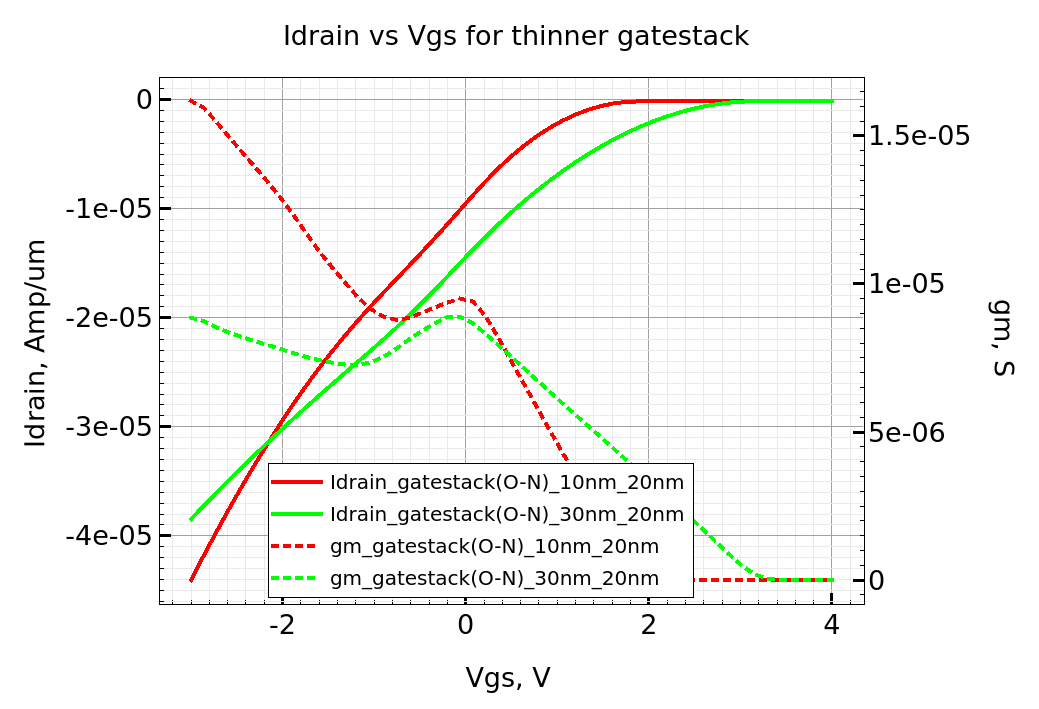}
    \caption{IV transfer curves and $g_m$ for different MOSFET channel implants (left) and gate oxide thicknesses (right).}
    \label{fig:gm_optimization}
\end{figure}

Increasing the buried-channel implant dose results in a larger channel conductivity and current density and therefore higher transconductance. Simulations, however, show marginal increase in $g_m$ at higher buried channel doses, possibly due to higher probability of scattering of the holes off the ions present in the channel, resulting in a decrease in the hole mobility.

Because, $g_m$ is directly proportional to the gate oxide capacitance, gate oxide thickness, on the other hand, strongly influences device performance. Thinner gate oxides improve electrostatic coupling between the gate and buried channel, resulting in a higher transconductance.

Figure \ref{fig:gm_optimization} shows the influence of buried channel implant (left) and gate oxide thickness (right) on $g_m$ of the device. 
The optimized design achieves transconductance values exceeding 16 $\mu$S compared with approximately 9 $\mu$S for the original structure.

\section{Gain Optimization} 
\label{sec:electronics}
\begin{figure}[hb!]
    \centering
   \includegraphics[width=.8\linewidth]{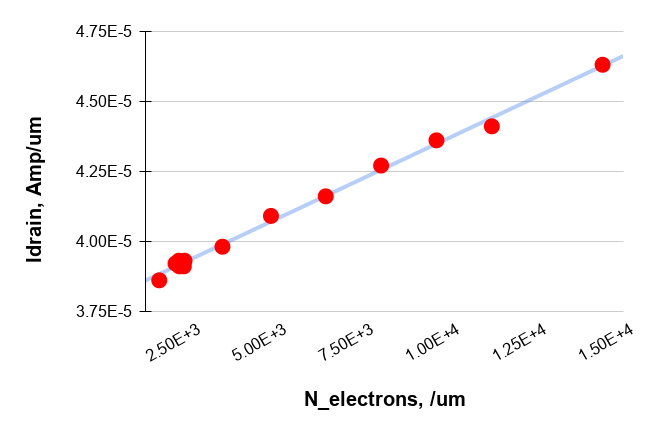}
    \caption{Drain current as a function of electron population in the internal gate. Gain is the slope of the fitted curve, which is around 511 pA/e for the baseline design.}
    \label{fig:gain_linearity}
\end{figure}

Detector gain ($G_q$) was determined from the change in drain current produced by individual electrons stored in the internal gate. We stored varying amount of electrons in the internal gate and determined the change in current from the baseline current density.
An example plot demonstrating the drain current as a function of number of electrons in the internal gate  is shown in Fig. \ref{fig:gain_linearity}. The measured current is fitted with a straight line, slope of which is the measurement of gain. We see even for a very large number of electrons, the gain is linear implying the large dynamic range these detector output stages can achieve. 

Here we discuss the influence of various design parameters on the gain of the device $-$

\subsection{Channel implant dose}
Increasing MOSFET channel implant dose produces a larger channel current and therefore stronger charge sensitivity. Simulations indicate a monotonic increase in gain with channel dose over the investigated range. It is to be noted that there is a certain limit in the channel current density that the device can withstand, therefore the channel implant dose can not be elevated indiscriminately. Beyond an implant dose of 7e12 /cm$^2$, the current density is too high for the safe operation of the amplifiers. 

\subsection{Trough implant energy}
The trough implant energy defines the vertical location of the internal-gate potential maximum and influences current density in the channel as well. Lower implant energies are found to improve gain. The fundamental quantity that defines the current density is the velocity of the holes in the channel for a fixed number of carriers. We found that the change in the hole velocity in the presence of electrons in the internal gate is higher for lower trough implant energies, yielding maximum change in the current. 

\subsection{Trough implant size}
Gain exhibits a non-monotonic dependence on trough size. Small troughs should yield lower capacitance to the channel than larger troughs. However, larger troughs influence a larger area of the channel for current modulation. For excessively small troughs, the capacitance might already be dominated by the fringing / parasitic capacitance at the expense of reduced charge coupling to the sensing channel. On the other hand, for excessively large trough sizes, a high capacitance lowers the gain. An intermediate trough width, therefore, produces maximum gain.
\begin{figure}[t!]
    \centering
   \includegraphics[width=1\linewidth]{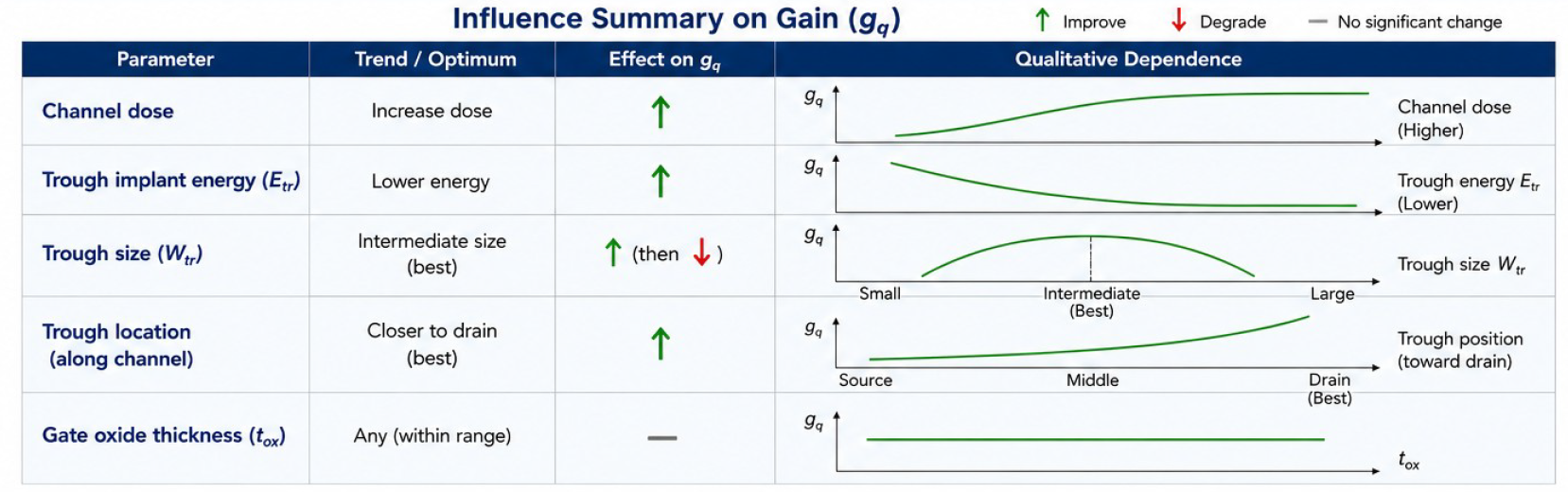}
    \caption{Influence of channel and trough design parameters on SiSeRO output-stage gain. Arrows indicate the direction of gain improvement and the sketches illustrate the observed qualitative trends. }
    \label{fig:gain_optimization}
\end{figure}

\subsection{Trough location}
The location of the trough beneath the MOSFET channel strongly affects gain. Simulations show a continuous increase in gain as the trough is moved from the source side toward the drain side of the channel, with maximum gain obtained for troughs positioned closest to the drain. When the stored electrons are located near the drain, their negative potential enhances the electric field in that region, resulting in increased hole acceleration toward the drain and, consequently, a higher drain current. In contrast, when the electrons are stored near the source, they increase the hole velocity only in the vicinity of the source. This effect decays rapidly along the channel, and beyond approximately 1 $\mu$m the hole velocity approaches that of the case without stored electrons, leading to a smaller increase in the drain current.

\subsection{Gate oxide thickness}
Within the investigated range, gate oxide thickness exhibits negligible influence on detector gain compared with channel and trough optimization.

Figure \ref{fig:gain_optimization} shows a qualitative trend in the gain improvement with the optimization of the process parameters. 
Overall, gain improves from approximately 511 pA/e in the baseline design to roughly 1100 pA/e in the optimized configuration.

\section{Noise Analysis}
\label{sec:results}
\begin{figure}[t!]
    \centering
   \includegraphics[trim={0cm 0cm 0cm 1.4cm}, clip=true,height=5.8cm, keepaspectratio]{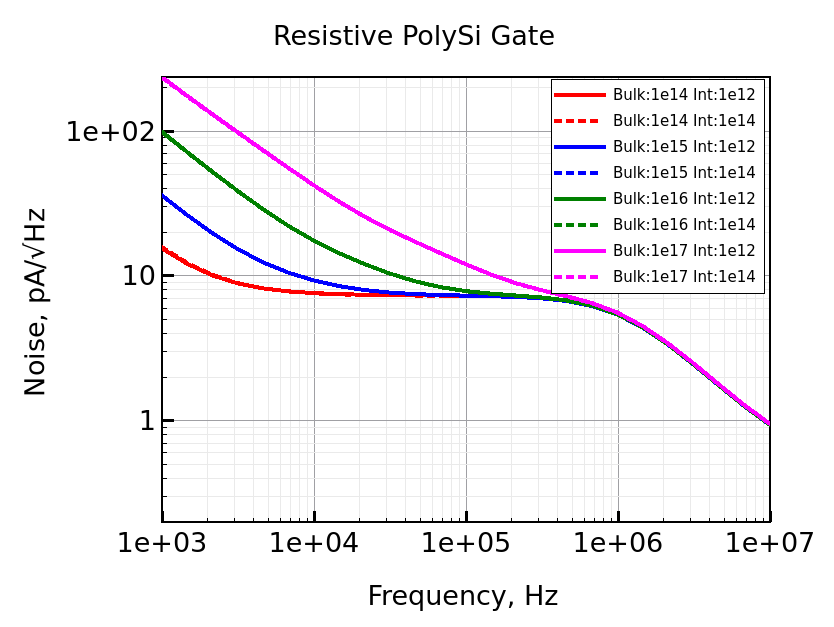}
   \includegraphics[trim={0cm 0cm 0cm 1.4cm}, clip=true,height=5.8cm, keepaspectratio]{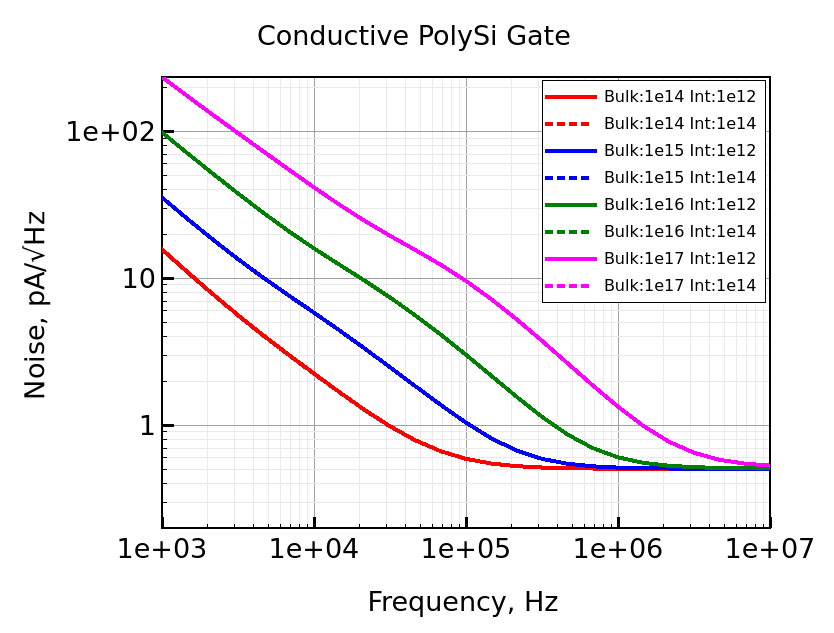}
    \caption{SiSeRO output stage noise spectral density for an undoped (left) and doped (right) polysilicon gate at 300 K. The different lines refer to varying silicon oxide interface trap density and bulk defect states.}
    \label{fig:noise_optimization}
\end{figure}

The noise mechanisms were investigated using frequency-domain ac simulations. In the observed noise spectral density of the first generation SiSeROs, we find a large 1/f noise with crossover frequency around 1 MHz. In order to reproduce the observed spectral density, in the simulations we included varying amount of interface traps and bulk defect states for low frequency correlated noise along with a thermal diffusion noise model. The simulated power spectral density is shown in the left plot of Fig. \ref{fig:noise_optimization}, consisting of a low-frequency 1/f component and a thermal-noise plateau. The crossover between these components occurs near 1 MHz when the bulk trap noise contribution is high. Interface trap noise contributes less strongly than bulk defects but remains important for achieving the lowest possible noise levels. The spectral shape follows the expected behavior for trap-induced fluctuations and suggests that defect reduction will be important for future device fabrications.

Thermal noise is found to be strongly influenced by polysilicon gate resistance. 
The gate resistance induced noise density in the channel can be modeled as 
\begin{equation}
    S~=~\frac{4KTR_{Gate}~g_m^2}{1~+~(\omega~R_{Gate}~C_{Gate})^2}
\end{equation}
At lower frequency, the plateau seen in the figure is proportional to $g_m$$^2$, so when current is small (V$_{SD}$ = 1V), noise plateau is low and with the increase in V$_{SD}$ or current, $g_m$ increases and eventually saturates to a higher plateau level. At higher frequencies above $F_{pole} ~=~ (1/2\pi R_{gate}C_{Gate})$, the gate resistance induced noise starts to fall as $1/f^2$ and settles down to a noise floor arising from the resistive channel of the MOSFET (see paper by Goo et al.\cite{goo2001} where they discuss similar effects from the substrate of the MOSFET). 

Simulations comparing undoped and heavily doped polysilicon gates show significant reduction in thermal noise when gate conductivity is increased (see the right plot of Fig. \ref{fig:noise_optimization}). We also inspected the influence of thermal annealing (rapid thermal annealing vs furnace annealing), dopant energy and dosage on the thermal noise, but found no significant dependence. 
The optimized structure achieves a simulated thermal-noise spectral density of approximately 0.5 $pA/\sqrt{Hz}$, which is a several factor lower than the undoped polysilicon based design.

We also inspected the noise for a thinner gate-stack, because for a thinner gate oxide with the increase in $g_m$, the MOSFET channel noise should also increase ($\propto\sqrt{g_m}$).  We find a slight increase (roughly by a factor of $\sqrt{2}$) in the thermal noise floor consistent with the factor of 2 increase in $g_m$.

\section{Implications for SiSeRO Active Pixel Sensors}
\begin{figure}[t!]
\centering
    \includegraphics[width=0.48\textwidth]{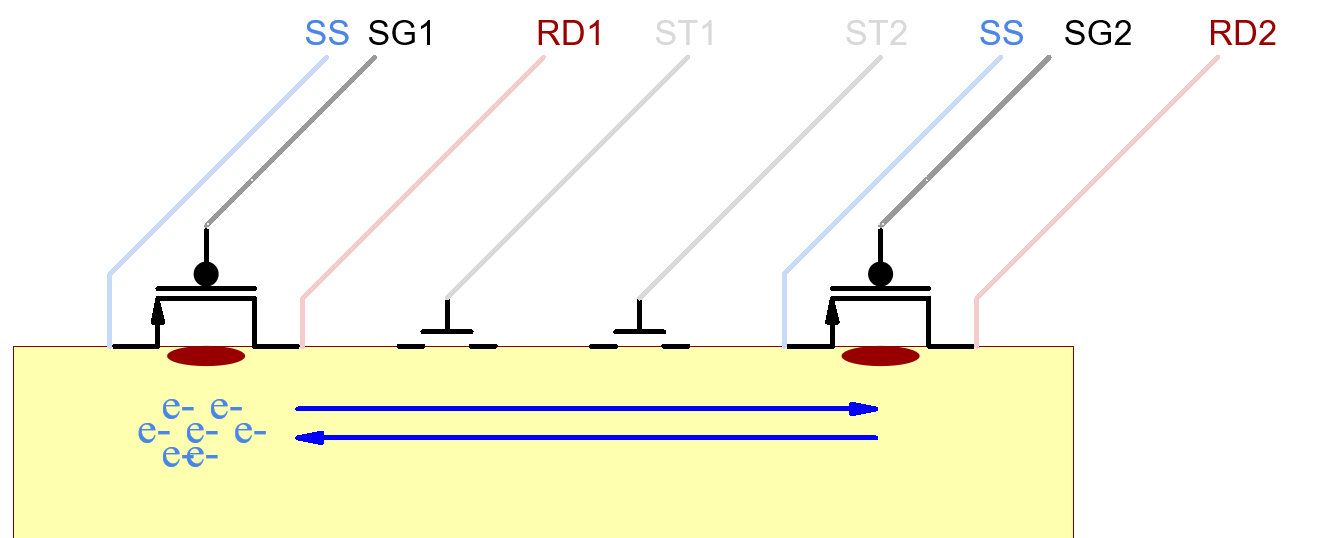}
    \includegraphics[width=0.5\textwidth]{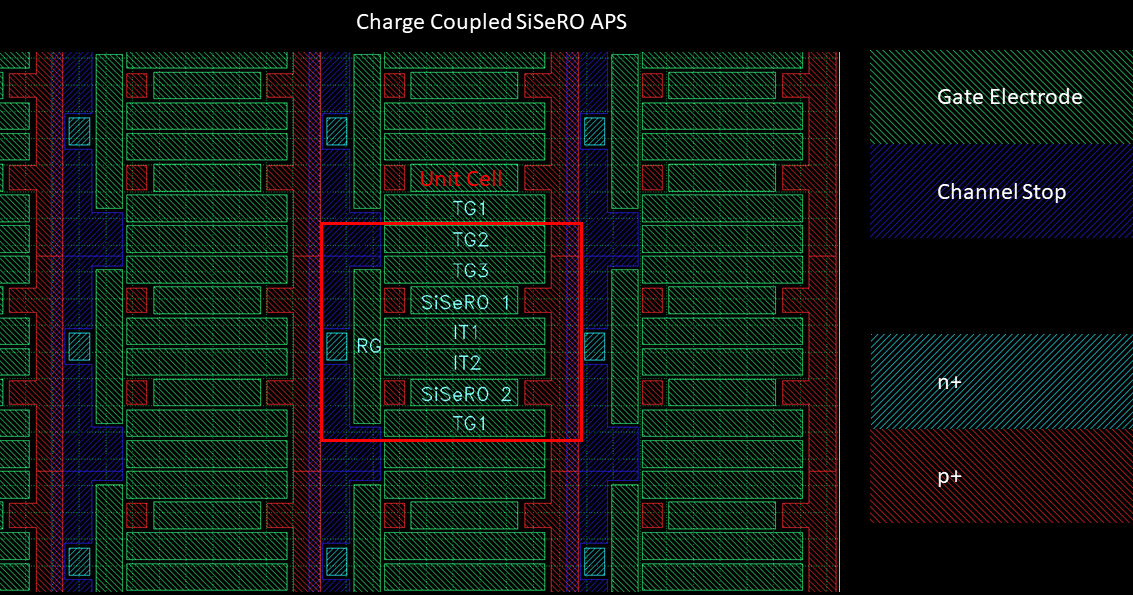}
    \caption{Left: Schematic of two SiSeRO transistors placed next to each other, with a dedicated transfer gate arrangement between them, allowing signal charge to be moved between the two amplifiers for repetitive measurements. Right: Preliminary layout of dual-SiSeRO pixel (denoted by the red square box). The pixel boundaries are established by a channel stop (blue) and barrier gate structure (TG1).}
\label{fig:SiSeRO_matrix}
\end{figure}

SiSeRO APS with RNDR provides the capability to read out larger detector arrays with low-noise at significantly higher speed. Such a device is also capable of region-of-interest readout in the same observation. With no macro charge transfer, APS devices are also in general more radiation hard and largely immune to radiation damage induced charge transfer inefficiency. 
Our APS concept design which is based on the RNDR optimized DEPFET detectors \cite{wolfel06}, employs two SiSeRO sensing nodes per pixel connected through a common electron-transfer channel. Signal charge can be transferred between the two nodes to enable RNDR operation through simultaneous measurement of signal and reference states. A cartoon schematic of the concept is shown in the left panel of Fig. \ref{fig:SiSeRO_matrix}, and the right image shows a preliminary layout for the structure. 
The two internal transfer gates (IT1, IT2) allow charge transfer between the two sense nodes. The pixel separation on the sides is achieved using channel stops like in conventional CCDs, whereas a barrier gate (TG1) provides separation on the top and bottom. Each pixel includes a reset gate and reset drain implant to empty the charge in the internal gate. 
We plan to incorporate the optimized output-stage design into the dual-SiSeRO active pixel sensor architecture.

\begin{figure}[t!]
\centering
    \includegraphics[height=7.3cm,keepaspectratio]{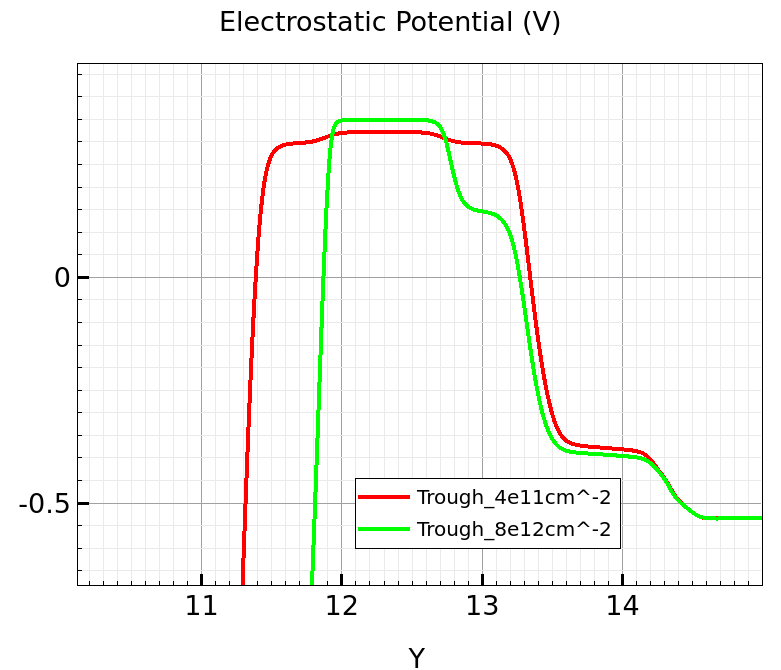}
    \caption{Localized and sharp peaked potential at the internal gate is obtained for elevated internal gate trough doses.}
\label{fig:SiSeRO_troughdose}
\end{figure}
Currently, we are developing three-dimensional TCAD process simulations for the dual-SiSeRO pixels. Since the SiSeROs share a common charge transfer channel, deeper trough potentials might be required in APS devices to prevent charge spilling during transfer operations. Simulations suggest that increased trough implant doses provide improved charge confinement and are therefore being investigated for future designs (see Fig. \ref{fig:SiSeRO_troughdose}).
\section{Summary and Future Work}\label{sec:summary}
We are working on two-dimensional TCAD simulations of the SiSeRO output-stage transistor in Synopsis Sentaurus to explore further improvement in gain and noise performance for future SiSeRO CCDs and active pixel sensor implementations.
Here we summarize the key findings $-$
\begin{itemize}
    \item Transconductance improves through increased buried-channel implant dose and reduced gate oxide thickness.
    \item Detector gain increases with channel implant dose and optimized trough geometry.
    \item Maximum gain is obtained for troughs located nearest the drain side of the channel.
    \item Bulk defect states dominate low-frequency noise.
    \item Polysilicon gate conductivity significantly influences thermal noise.
\end{itemize}

The optimized design increases detector gain from approximately 511 pA/e to 1100 pA/e while simultaneously improving transconductance and reducing noise. Future work will include three-dimensional TCAD simulations, optimization of channel width and explore improved isolation of the sense node from the surrounding structures. We plan to develop process simulations for SiSeRO APS devices and validate the dual-SiSeRO RNDR operation. These simulations will play an important role and form the base for the design and development of future active pixel SiSeRO matrix, representing the next step toward large-format single-electron-sensitive fast imaging detectors.

\acknowledgments 
This work was supported by the NASA Astrophysics Research and Analysis (APRA) program under contract number 80NSSC25K7957 and by the NASA Strategic Astrophysics Technology (SAT) program under grant number 80NSSC23K0211.


\end{document}